\documentstyle[12pt]{article}
\def\btabl{\begin{table}}   \def\etabl{\end{table}}
\def\bea{\begin{eqnarray}}   \def\eea{\end{eqnarray}}
\def\bnn{\begin{eqnarray*}}   \def\enn{\end{eqnarray*}}
\def\beq{\begin{equation}}   \def\eeq{\end{equation}}  
\def\btabu{\begin{tabular}}   \def\etabu{\end{tabular}}
\def\bec{\begin{displaymath}} \def\eec{\end{displaymath}}
\def\nn{\nonumber}
\def\eqref#1{(\ref{#1})}
\begin{document}

\begin{titlepage}
\begin{flushright} FANSE-96/09\\LPTHE Orsay-96/24\\ hep-ph/9604273
\end{flushright}

\begin{centering}

{\large \bf Relativistic Quantum Scattering of High Energy Fermions in the 
Presence of Phase Transition}
\bigskip

{\large Jos\'e Rodr\'{\i}guez-Quintero and Manuel Lozano}

\bigskip

\noindent
Departamento de F\'{\i}sica At\'omica, Molecular y Nuclear, Universidad de 
Sevilla, Spain\footnote{Work partially supported by Spanish CICYT, project 
PB 92-0663}\\
\bigskip

{\large Olivier P\`{e}ne}

\bigskip

\noindent
LPTHE, F 91405 Orsay, France\footnote{Laboratoire associ\'e au Centre National 
de la Recherche Scientifique, URA  D0063.}\\
\end{centering}

\begin{abstract}

We study the high energy behaviour of fermions hitting a general wall caused
by a first-order phase transition. The wall profile is introduced through 
a general analytic function. The reflection coefficient is computed in the
high energy limit and expressed in terms of the poles of the wall profile
function. It is shown that the leading singularity gives the high energy
behaviour.

\bigskip
\bigskip

\noindent PACS. \ :

\begin{itemize}

\item{11.10Q-}Field theory, relativistic wave equations.
\item{11.80F-}Relativistic scattering theory, approximations.
\item{11.30Q-}Symmetry and conservation laws, spontaneous symmetry breaking.

\end{itemize}
\end{abstract}
\medskip

\end{titlepage}

Much work have been devoted to the problem of transmission and reflection
of relativistic fermions through a wall separating two phases of different
symmetry properties. The main effort concentrates in developing the idea \cite{kuzmin} that
the baryon asymmetry of the Universe might have been produced if the  
cosmological electroweak ($SU(2) \times U(1)$) phase transition has been  
of first order. In these works the phase transition is described in terms of 
bubbles of "true" vacuum with an inner expectation value of the Higgs field
$v \neq 0$, i.e. a spontaneously broken symmetry phase, appearing and 
expanding in the preexisting "false" vacuum with $v=0$, i.e. a symmetric
phase. 

In this scenario the quarks/antiquarks hitting the wall from the
unbroken phase are reflected or transmitted. The point to elucidate is
whether there exists a CP-asymmetry that produces a different reflection
and transmission probability for quarks and antiquarks in
order to explain, via the standard model baryon number anomaly \cite{thooft}, 
\beq
\partial_\mu J^\mu_B=\partial_\mu J^\mu_L= N_f (\frac {g^2}
{32 \pi^2}W \widetilde W -\frac{g'^2}{32 \pi^2} Y\widetilde Y).\label{anomalie}\eeq
the correct baryon asymmetry of the Universe. 
In the 
physical conditions of the early universe the fermions moving through the
bubble wall will interact not only with the wall but also with the particles
in the surrounding plasma, thus we have a transport problem. 
This transport plays an essential in the so-called ``charge transport" mechanism \cite{Nel92}, in which the action of the baryon anomaly happens at a distance from the bubble wall. This diffusion
problem is very complicated and involves solving the Fokker-Planck equation taking into account CP violation and baryon anomaly. A useful simplifying assumption is to decompose the process into two steps, one describing the
production of the CP asymmetry when the quarks/antiquarks are reflected on the wall, the second describing the transport and the eventual transformation of the CP asymmetry into a baryon asymmetry via the baryon number anomaly \eqref{anomalie}. \par

One expects that the diffusion corrections are relatively minor ones to
the first step, i.e. to the scattering from the wall, although effects of the surrounding plasma 
are incorporated by mean of introducing the Higgs field effective potential 
that takes into account the temperature of a thermal bath. The structure of
the wall depends on this effective potential and its knowledge is obviously
necessary in order to compute the reflection and transmission coefficients.
However, the profile obtained by solving the equation of motion is rather
complex and depends on many coupling constants 
\cite{Nel92}. At this point, two lines for simplifying the problem have been
followed. The first is to describe the profile of the bubble wall by an
analytical function that simulates the dynamics of the phase transition and
then to treat the scattering in an approximate way[1-3]. 
The second approximates the wall profile by a step function
i.e. a sudden jump from one phase to the other, the  thin wall approximation \cite{shapo},
\cite{Gav94}. this extreme simplification  allows to compute in an exact way the Feynman fermion
propagator in the presence of a wall \cite{Gav94}. \par

Our aim in the present paper is to study
the general problem of fermions hitting a wall in order to connect the analytic
properties of the profile function to the behaviour of the fermions. In 
general, this question is very difficult, but interesting conclusions about 
this relation between the profile and the behaviour of fermions in the high 
energy limit can be obtained.
We develop a general method to calculate the reflection and transmission
coefficients of high energy fermions hitting a wall, establishing a relationship
between the poles of the profile function and these coefficients.
The quantum corrections to the expected classical behaviour are obtained.
Apart from the clear theoretical and purely formal interest of this result, 
cosmological implications and possible applications of the formalism to systems of
relativistic fermions undergoing a phase transition should be considered. Some
implications in CP-violating process are studied in \cite{Rod96}. Two other 
examples of application fields can be condensed matter under extreme external 
conditions or certain stages of the quark-gluon plasma formation process. 

As usual in this kind of work, we formulate the problem in the rest frame
of a wall, parallel to the 
$x-y$ plane and normal to the z-axis, characterized by a general profile. In order to calculate the reflection 
coefficient we need only the plane wave solutions of the Dirac equation for 
particles moving along the z-axis. In any case, for other incoming directions
we can perform an appropriate Lorentz boost parallel to the $x-y$ plane and reduce the problem to the 
latter. The phase transition is incorporated into the Dirac equation by including a position dependent
 mass term which varies inside a certain region designated
as the domain wall and takes two different constant values in the two outer sides of the wall.
Following Nelson et al\cite{Nel92}, we work in the chiral basis, conveniently reordered to obtain 
$\gamma^5= \pmatrix{\sigma_3 &0 \cr 0 &- \sigma_3 \cr}$, and factor
the Dirac operator into $2 \times 2$ blocks. Thus the Dirac equation can be 
expressed as

\beq
\pmatrix{
i \partial_z + i \partial_t &- m(z) &0 & 0 \cr
m^*(z) &i \partial_z - i \partial_t &0 &0 \cr
0 &0 &i\partial_z + i \partial_t &- m^*(z) \cr
0 &0 &m(z) &i \partial_z - i \partial_t \cr
} \Psi = 0. \label{1} 
\eeq

\noindent With the following Ansatz for solutions with positive 
energy $E$

\beq
\Psi =  {\psi_I \choose \psi_{II}} e^{-iEt}
 \qquad \hbox{with} \quad \matrix{\psi_I = {\psi_1 \choose \psi_2}  \cr
\psi_{II} = {\psi_3 \choose \psi_4} \cr } {\label{2}}
\eeq
where $\psi_1$ and $\psi_4$ are eigenspinors of the chirality operator, $\gamma_5$, 
for the eigenvalue $+1$ and $\psi_2$ and $\psi_3$ for $-1$, we obtain
\bea
&&\left ( i \partial_z + Q(z) \right ) \psi_I = 0 \nn \\
&&\left ( i \partial_z + \overline{Q}(z) \right ) \psi_{II} = 0 \label{3} 
\eea
\noindent{ being}

\beq
{\em Q}(z) = \left( \begin{array}{cc} E & -m(z) \\ m(z)^* & -E \end{array}
\right) \;\;\;\; \mbox{and} \;\;\;\; {\overline{\em Q}(z)} = \left( 
\begin{array}{cc} E & -m(z)^* \\ m(z) & -E \end{array} \right);
\label{QdeZ}
\eeq

\noindent  where the mass function can be considered as complex in order to 
incorporate CP-violating process in the formalism. In this work we are not
interested in CP-violation but in a general approach of the problem. Thus the
mass function will be assumed as real and consequently ${\em Q}(z)$
and ${\overline{\em Q}(z)}$ will be identical matrices.  
 The solution for 
the first equation of (\ref{3}) can be written as follows

\beq
\psi_I(z) = {\em P}e^{i\int^z_{z_0}d\tau {\em Q}(\tau)} \left( \begin{array}{c}
\psi_1(z_0) \\ \psi_2(z_0) \end{array} \right)
\label{5}
\eeq

\noindent{and analogously for the second one}

\beq
\psi_{II}(z) = {\em P}e^{i\int^z_{z_0}d\tau {\overline{\em Q}(\tau)}} \left( 
\begin{array}{c} \psi_3(z_0) \\ \psi_4(z_0) \end{array} \right),
\label{6}
\eeq

\noindent where {\em P} indicates a path ordered product and $\tau$ is the 
position variable along the z-axis. Nevertheless, it is obvious that for a
real mass function the quantity

\beq
\Omega(z,z_0)={\em P}e^{i\int^z_{z_0}d\tau {\em Q}(\tau)}   \label{Omega}
\eeq

\noindent{in (\ref{5}) is the same than ${\overline \Omega(z,z_0)}={\em P}
e^{i\int^z_{z_0}d\tau {\overline{\em Q}(\tau)}}$ in (\ref{6}).
In consequence, the task is to evaluate equation (\ref{Omega}).

Using the usual Pauli's matrices, $Q(\tau)$ can be expressed as

\beq
{\em Q}(\tau) = \sigma_3 E - i \sigma_2 m(\tau).
\label{10}
\eeq

\noindent We consider now

\beq
m(\tau)=m_0f(\tau),
\eeq

\noindent where $f(\tau)$ is a certain function which describes the structure 
of the domain wall, the profile wall function. 
The asymptotic conditions $f(+\infty)=1$ and $f(-\infty)=0$ are required, 
$f(\tau)-\theta(\tau)$ decreasing exponentially when $\tau \to \pm\infty$. 
it is also assumed that $f(\tau)=O(1)$. The profile function
will be considered as an analytic function in the real axis, therefore a 
profile which is constant outside a certain finite region, the domain wall, 
cannot be described. This last kind of wall profiles are studied in detail
in \cite{Rod96}. If the profile function is analytic, the wall, defined
as the region where the mass varies with the position, extends formally from
$\tau=-\infty$ to $\tau=+\infty$. Nevertheless the domain wall will be 
characterized by an effective thickness, $\sigma$. The particular criterion 
considered in order to define this parameter is not important; for instance, it 
can be established by taking into account that $|f(\tau)-1|<0.1$ for 
$\tau > \sigma$ and $|f(\tau)|<0.1$ for $\tau < -\sigma$. 
However, in general we can write 

\beq
f(\tau)=F \left( {\tau \over \sigma} \right) \ .
\label{FdeTau}
\eeq

\noindent A characteristic energy, $m_0$, and a characteristic length, $1/m_0$, 
will be used to obtain dimensionless quantities in what follows. 
By considering a certain path partition $(z_0,z_1,......,z_{N-1},z_N,z)$, we can
write:

\beq
{\cal P} \ e^{i\int_{z_0}^z d\tau \ Q(\tau )} = {\cal P} \ 
e^{i \int_{z_N}^zd \tau \
Q(\tau )} {\cal P} \ e^{i\int_{z_{N-1}}^{z_N} d\tau \ Q(\tau )} ... \ 
{\cal P} \ e^{i\int_{z_0}^{z_1} d\tau \ Q(\tau )} \ \ \ . \label{A2}
\eeq

\noindent As shown in the appendix, we find for small $\Delta_j$, where 
$\Delta_j = z_{j+1}-z_j$, that 

\beq
{\cal P} \ e^{i \int_{z_j}^{z_{j+1}} d \tau \ Q(\tau )} = e^{i\sigma_3 p_j
\Delta_j} +{1 \over E} \sigma_2 f_j\sin (p_j \Delta_j) \ \ . 
\label{A6} 
\eeq 

\noindent where $p_j = + (E^2 - f_j^2)^{1/2}$ and $f_j$ is defined as

\beq
f_j = {1\over \Delta_j}\int^{z_{j+1}}_{z_j}d\tau f(\tau),
\eeq

\noindent e.g. as the average value of the integral.}

Thus, by substituting the path order products of the right-hand side in 
(\ref{A2}) by (\ref{A6}), multiplying and reordering the terms, we obtain

\bea
&&\Omega (z, z_0) = e^{i\sigma_3 \sum\limits_{j=0}^N p_j \Delta_j} +  \nn \\ 
&&+ {1 \over E} \sigma_2 \sum\limits_{k=0}^N \sin (p_k \Delta_k)f_k \ 
e^{-i\sigma_3\sum\limits_{j=k+1}^N p_j \Delta_j} \ 
e^{i\sigma_3 \sum\limits_{j=0}^{k-1} p_j \Delta_j} 
+ O\left[\left({1\over E}\right)^2\right] ; 
\eea 

\noindent where we have used $e^BA=Ae^{-B}$ when $\{ A,B \}=0$.
If the path partition is now considered to be thiner and thiner, or in other
words, if the limit $\Delta_j \to 0$ is taken for all $j$, then we can replace
$\sum\limits_j\Delta_j \to \int d\tau$. The definition of $f_j$ as an integral
average value leads to the following substitutions $f_j \to f(\tau)$ and $p_j
\to p(\tau)$, where $p(\tau)=+\left( E^2-\left[f(\tau)\right]^2\right)^{1/2}$,
provided that $f(\tau)$ is continuous.
In this way, we obtain

\bea
\Omega (z, z_0) = \left( 1+ \left( {1 \over E}\right)\sigma_2 \int_{z_0}^z 
d\tau \ p(\tau ) \ f(\tau ) \ e^{-2i\sigma_3\int_{\tau}^z d\xi p(\xi)} \right.
\nn \\
+ \left. O\left[\left({1\over E}\right)^2\right] \right) 
e^{i\sigma_3 \int_{z_0}^z d \tau \ p(\tau )} \ . \label{11} 
\eea

We must here stress that $\Omega(z,z_0)$ has been expanded in powers of
$1/E$, but not in a strict sens because the coefficients are also depending on 
$E$. As a consequence of this fact, whether the series for $\Omega(z,z_0)$ which 
is truncated in order to give (\ref{11}) is an asymptotic one or not must be 
carefully elucidated. This point is treated in detail in \cite{Rod96} and we
will return to this question below.
If we define the quantity $a=2\sigma E$ and expand all the terms which depend
on $E$ in (\ref{11}) in powers of $1/E$, by assuming $a$ constant, we obtain

\beq
\Omega (\lambda\sigma,-\lambda\sigma) = 
\left \{ 1 + {1 \over E} \sigma_2 {a \lambda \over 2} \int_{-1}^1 dx \
F(\lambda x) \ e^{-i \sigma_3 a \lambda (1 - x)} + 
O \left [ \left ( {1 \over E} \right )^2 \right ] \right
\} e^{i\sigma_3 \phi} \ \ \ ; \label{36}
\eeq

\noindent where $z=\lambda \sigma$ and $z_0=-\lambda \sigma$, with $\lambda \gg
1$. The function $F(\tau)$ has been introduced in (\ref{FdeTau}) and the 
integration variable is changed to $x={\tau \over \lambda \sigma}$. 
The quantity $\phi$ is defined 
as "{\cal classical path}", $\int^z_{z_0}d\tau p(\tau)=\lambda a + O\left[ 
\left({1\over E} \right)^2 \right]$. Equation (\ref{36}) can be written in a 
matrix way

\bea
\Omega(\lambda \sigma,-\lambda \sigma)=
\left(  \begin{array}{cc} 
\omega_{11} & \omega_{12}  \\ 
\omega_{21} & \omega_{22}
\end{array} \right).
\label{Momega}
\eea

\noindent Neglecting terms of the order $o\left( 1/E \right)$ we get

\bea
&&\omega_{11}=e^{i\lambda a}=\omega_{22}^*,  \nn  \\
&&\omega_{12}=-{ia\lambda \over 2}{1\over E} {\cal I}(a)^*=\omega_{21}^*;
\label{omega}
\eea

\noindent where

\beq
{\cal I}(a) = \int^1_{-1}dxe^{i\lambda ax}F(\lambda x).
\label{IdeA}
\eeq

By considering the requirements imposed to the profile wall function, $f(\tau)$,
the following approximations are performed: $m(\tau)=0$ for $\tau < 
-\lambda \sigma$ and $m(\tau)=m_0$ for $\tau > \lambda \sigma$; i.e. 
$F(\lambda x)=0,1$ for $x<-1$ and $x>1$, respectively.
Thus, taking into account the chirality eigenvalues of the eigenspinors
$\psi_1$, $\psi_2$, $\psi_3$ and $\psi_4$, we have that $\psi_1$ and $\psi_2$ 
correspond to right-moving right-handed particles and left-moving left-handed 
particles respectively, and $\psi_3$ and $\psi_4$ to right-moving left-handed 
and left-moving right-handed particles, in the region where $x<-1$.  
Concerning the region where $x>1$, $\Omega(\lambda \sigma x,\lambda \sigma x)$
can be immediately diagonalized in order to identify the right-moving and the
left-moving flux of particles, since the mass is constant. Thus, by requiring
that the left-moving flux is zero in the latter region, i.e. by imposing that
we only have a transmitted flux in this region, a left-handed reflected flux
is obtained from the right-handed one and vice versa\footnote{Indeed, by conservation of angular momentum $J_z$ the helicity is flipped in the reflection process.}. The equivalence between
the equations for $\psi_I$ and for $\psi_{II}$ in (\ref{3}), as a consequence 
of the result $Q(z)={\overline Q(z)}$ for the real mass case, implies that 
there is no asymmetry in the reflection of incident right-handed and left-handed
fermions. In consequence, needless to distinguish $\psi_I$ and $\psi_{II}$. Only 
when an imaginary mass term, which introduces CP-violating effects, is 
incorporated, differences between the behaviour of right-handed and 
left-handed fermions may exist in the reflection 
process[1,3,5]. The fact that CP invariance boils down to an equality between both helicities is a consequence of CPT invariance \cite{Gav94}.

In this way, following the method above described, which was originally 
introduced by Nelson {\it et al} for a numerical computation of the reflection
coefficient[6,7], we use that

\beq
\psi(z)=\Omega(z,z_0)\psi(z_0),
\label{Ome1}
\eeq

\noindent where

\bea
\psi(z)= {\em D}
\left(  \begin{array}{c}  T e^{ipz} \\  0  \end{array}  \right)
& \mbox{and} & 
\psi(-z)=
\left(  \begin{array}{c}  e^{iEz_0} \\ R e^{-iEz_0}   \end{array}  \right).
\label{Ome2}
\eea

\noindent ${\em D}$ diagonalizes the matrix $Q$ in the region where the mass is
constant. From eq. (\ref{Ome1}), (\ref{Ome2}) we obtain for the reflection 
coefficient

\beq
R(E) = - e^{2iEz_0} 
{\omega_{12}^* - (E - p) \omega_{11} \over \omega_{11}^*
- (E - p) \omega_{12}} \ \ .
\label{RdeE}
\eeq

\noindent with $p=+\sqrt{E^2-1}$. In our case, for the high energy limit, we
have

\beq
R(E)={1\over 2E}\left( e^{i\lambda a}-i\lambda a {\cal I}(a) \right) 
+ o\left[ \left( {1\over E} \right)^2 \right] \ .
\label{RFdeE}
\eeq

Now, we must compute the integral which gives the function ${\cal I}(a)$ in
eq. (\ref{IdeA}).
We introduce a certain convergence factor, $e^{-\varepsilon x}$, where the
parameter $\varepsilon$ is taken small enough to be sure that $F(\lambda x)$
decreases faster than this factor when $x \to -\infty$, in such a way that
the integration over the real axis, from $-\infty$ to $\infty$, is well defined.
Thus, by cutting the integration domain into three intervals, $(-\infty,-1)$,
$(-1,1)$ and $(1,\infty)$, we can write

\beq
{\cal I}(a)=\lim_{\varepsilon \rightarrow 0^+}
\left(
\left\{ \int_{-\infty}^{\infty}-\int_1^{\infty}-\int_{-\infty}^{-1} \right\}
dxe^{(i\lambda a-\varepsilon)x}F(\lambda x)
\right) \ .
\label{IFdeE}
\eeq

\noindent  where the integrals in the regions $(-\infty,-1)$ and $(1,+\infty)$ 
can be trivially solved. Indeed, as indicated above, we approximate 
$F(\lambda x)=0,1$ for $x<-1$ and $x>1$, respectively. Therefore we have

\bea
&&\lim_{\varepsilon \to 0^+} \int_{-\infty}^{-1} dx \ F(\lambda x) \ 
e^{(i\lambda a - \varepsilon )x} = 0 \ ,  \nn  \\
&&\lim_{\varepsilon \to 0^+} \int_{+1}^{+\infty} dx \ F(\lambda x) \ 
e^{(i\lambda a - \varepsilon )x} = {i\over \lambda a} e^{i\lambda a} \ .
\label{IIs}
\eea

Concerning the integral over all the real axis, the Cauchy theorem can be 
applied in order to perform this integration.
If the Laurent expansion for $F(z)$ in the pole of order $\nu_j$, $z = z_j$, is
$\sum\limits_{n=-\nu}^{+ \infty} b_n^j (z - z_j)^n$ the following result can be 
obtained

\beq
\lim_{\varepsilon \to 0^+} \int_{-\infty}^{+\infty} dx \ F(\lambda x) \ 
e^{(i\lambda a - \varepsilon )x} = {2 \pi i \over \lambda} 
\sum_{j=1}^N e^{-a y_j} \ e^{iax_j} 
\sum_{n=1}^{\nu_j} b_{-n}^j {(i a )^{n-1} \over (n - 1) !} \ ; 
\label{Ipole}
\eeq

\noindent where $z_1 = x_1 + iy_1$, $z_2 = x_2 + i y_2$, ..., 
$z_N = x_N + iy_N$, are all the poles of $F(z)$ with positive imaginary part 
we have picked when the integration contour is adequately closed. Since the 
profile function is analytic on the real axis, it is obvious that $y_j \not= 0$ 
and therefore we obtain an exponential dependence on $a$. 

It can be shown from eqs. (\ref{RFdeE}), (\ref{IFdeE}), (\ref{IIs})  and 
(\ref{Ipole}) that

\beq
R(E) = {\pi a \over E} \sum_{j=1}^N e^{-a y_j} \
e^{iax_j} \sum_{n=1}^{\nu_j} b_{-n}^j {(i a )^{n-1} \over (n - 1) !} 
+ o\left[ \left( {1\over E} \right)^2 \right] \ . 
\label{Rpole}
\eeq

Obviously, $R(E)$ defined in eq. (\ref{RdeE}) and given by eq. (\ref{RFdeE}) is 
the reflection coefficient generated by a profile function which is zero in the 
region $(-\infty , -\lambda \sigma)$ and constant in the region 
$(\lambda \sigma , \infty)$.
Nevertheless, $f(\tau)$ is an analytic function in the real axis verifying that
$f(\tau)-\Theta(\tau)$ decreases exponentially when $\tau \to \pm \infty$, in 
such a way that the wall profile is well defined over all the real axis. In 
fact, we are interested in the situation where the wall profile extends from
$-\infty$ to $\infty$, although the former requirement is technically necessary 
in order to identify the incident, reflected and transmitted waves. On the 
other hand, we cannot take directly the limit when 
$\lambda \to \infty$\cite{Rod96}. 
This problem is solved by considering $\lambda$ as large enough to neglect the 
effects due to the evolution of the wave outside the region 
$(-\lambda \sigma , \lambda \sigma)$.
The question is studied in detail way in ref. \cite{Rod96} and the result is, 
as intuitively expected, that 
in the limit where we take for the wall profile an analytic function from 
$-\infty$ to $+\infty$ one simply has to consider in \eqref{IFdeE} the first 
integral, from $-\infty$ to $+\infty$, and that consequently (\ref{Rpole}) 
{\it gives the reflection coefficient} for the wall from 
$-\infty$ to $+\infty$. 
The effect of the mentioned technical assumption (the substitution
of the true profile function by a step function outside the interval), is that
 ${\cal I}(a)$ is defined as an integral over de domain $(-1,1)$. It can be 
bounded as seen in (\ref{IFdeE}) when this function, ${\cal I}(a)$, is expressed
through the integral over all the real axis and the two other integrals given 
by eq. (\ref{IIs}). It is worth to stress that indeed $R(E)$ in eq. 
(\ref{Rpole}) does not depend on $\lambda$, as expected for the reflection 
coefficient for the wall from $-\infty$ to $+\infty$. 

In (\ref{RFdeE}) the leading term for a series of powers in $1/E$ has been 
written. Nevertheless, we can be sure that it gives the leading term only if
$a$ is a constant. If the parameter which is considered constant is $\sigma$,
then we must return to the problem of the asymptotic properties of the 
series we use. As above mentioned, this question is discussed in \cite{Rod96},
where we conclude that the right-hand side of (\ref{RFdeE}) is the leading term
of the reflection coefficient for high energy when $\sigma \ll 1$, even if 
$a=2 \sigma E \gg 1$.
In this way, by considering the dependence on $E$ for the reflection coefficient
in the high energy limit, taking $\sigma$ as constant, we have

\beq
R(E) = 2 \pi \sigma \sum_{j=1}^N e^{-2E\sigma y_j} \
e^{2iE\sigma x_j} \sum_{n=1}^{\nu_j} b_{-n}^j 
{(2iE \sigma )^{n-1} \over (n - 1) !} \ . 
\label{RFpole}
\eeq

\noindent Thus, the behaviour of the high energy fermions, i.e. the high 
energy component of a wave packet, hitting the wall is completely determined 
by the complex 
singularities of the wall profile function. Furthermore, if we are in such a
range of energy that $\sigma E \gg 1$, the leading contribution is exponentially decreasing with the energy, the coefficient of the exponential being given by 
the pole of $f(z)$ with the smallest imaginary part, i.e. the closest to the 
real axis. Calling $z_k = x_k + i y_k$ this prevailing pole we obtain

\beq
R(E) = 2 \pi \sigma e^{-2E\sigma y_k} 
\ b_{-\nu_k}^k \ 
{(2i E \sigma )^{\nu_k - 1}\over ( \nu_k - 1) !} \ 
e^{2iE\sigma x_k} \ . \label{RLpole}
\eeq

\noindent The dependence in $\sigma$ in \eqref{RLpole} is simply an artifact of 
our change of variables \eqref{FdeTau}. Indeed it can be inmediately seen that
if we consider the Laurent expansion, 
$\sum\limits_{n=-\nu}^{+ \infty} b_n^{'j} (z - z_j)^n$, for $f(z)$ instead 
$F(z)$, in the pole of the order $\nu_j$, $z=z'_j$, we have

\bea
&& z'_j = \sigma z_j \ , \nn \\
&& b^{'j}_n = {b_n\over \sigma^n} \ ;
\label{ZandB}
\eea

\noindent
where, by using \eqref{FdeTau}, the poles and the coefficients of the Laurent
expansion for $F(z)$ and $f(z)$ have been related. Thus, the final result 
can be expressed in terms of the singularities of $f(z)$ by replacing 
\eqref{ZandB} in eq. \eqref{RFpole} and \eqref{RLpole},

\bea
R(E) = 2 \pi \sum_{j=1}^N e^{-2E y'_j} \
e^{2iE\sigma x'_j} \sum_{n=1}^{\nu_j} b_{-n}^{'j} 
{(2iE)^{n-1} \over (n - 1) !} \nn \\
= 2 \pi e^{-2E y'_k} 
\ b_{-\nu_k}^{'k} \ 
{(2i E)^{\nu_k - 1}\over ( \nu_k - 1) !} \ 
e^{2iE x'_k} \ ;
\label{FinEq}
\eea

\noindent where the lower expression is valid in such a rang of energy that
$e^{-2E(y'_k-y'_h)}$ can be neglected, $y'_k$ and $y'_h$ being the imaginary 
parts of the two closer poles to the real axis (this requirement is analogous 
to the former $\sigma E \gg 1$). The dependence on $\sigma$ has disappeared in 
\eqref{FinEq}, as expected.

The results we present can be checked by using the particular analytic solution 
obtained in ref. \cite{Fun94}, \cite{Aya94} for the following wall profile

\beq
f(\tau ) = {1 + {\tanh}(\tau / \sigma ) \over 2} \ . \label{46}
\eeq

\noindent For this particular profile function, it follows from (\ref{RFpole}) 
by computing the sum for the infinite series of poles it presents

\beq
R(E) = {\pi \sigma \over 2 \sinh (\pi E \sigma )} \ , \label{47}
\eeq

\noindent which agrees with the result of reference \cite{Fun94} and 
\cite{Aya94} in the high energy limit for $\sigma \ll 1$. \par

With this positive check, we conclude that (\ref{RFpole}) in general 
and (\ref{RLpole}) for $E \sigma$ large give the high energy behaviour of the 
reflection coefficient for analytic profile functions in the real axis. The
singularities of the profile function determine the behaviour of the high 
energy fermions in the way shown by eq. (\ref{RFpole}), and if the energy is
larger enough than the inverse of the characteristic wall thickness, it is the
closest pole to the real axis that gives the high energy behaviour through eq.
(\ref{RLpole}).

\section*{Acknowledgements.}
We acknowledge Jean-Claude Raynal for early inspiring discussions and in 
particular for mentioning that the singularities of the wall profile functions 
might play an important role.

\bigskip

\appendix

\section{Appendix}

We take $Q(\tau )$ as constant in the intervals $(z_j , z_{j+1})$, defining
for each one $f_j$ as the following integral average value

\beq
f_j = {1 \over \Delta_j} \int_{z_j}^{z_{j+1}} d \tau \ f(\tau ) \ \ \ , 
\eeq

\noindent where $\Delta_j = z_{j+1} - z_j$. By assuming that $\Delta_j$ is small
(in fact, the limit $\Delta_j \to 0$ is considered for the result obtained in
this appendix), it can be written in good approximation by using (\ref{10})

\beq
{\cal P} \ e^{i\int_{z_j}^{z_{j+1}} d\tau \ Q(\tau )} = 
e^{(i\sigma_3E + \sigma_2f_j)\Delta_j} \ . 
\label{AA4}
\eeq

\noindent Taking into account that for two operators verifying $\{A,B\} = 0$ 
and $A^2 = B^2 = 1$, it can be easily proven the following result 

\beq
e^{\alpha A + \beta B} = \cosh \left [ \left ( \alpha^2 + \beta^2 \right )^{1/2} \right
] + {\alpha A + \beta B \over \left ( \alpha^2 + \beta^2 \right )^{1/2}} \sinh \left [
\left ( \alpha^2 + \beta^2 \right )^{1/2} \right ] \ \ \ , \label{AA5}
\eeq

\noindent and from (\ref{AA4})

\bea
{\cal P} \ e^{i \int_{z_j}^{z_{j+1}} d \tau \ Q(\tau )} = 
e^{i\sigma_3 p_j\Delta_j} + {1 \over E} \sigma_2 f_j \sin (p_j \Delta_j) 
+ O\left [\left ( {1 \over E} \right )^2 \right ]  \ . 
\nn   (\ref{A6})
\eea 

\noindent where $p_j = + (E^2 - f_j^2)^{1/2}$. In order to obtain (\ref{A6}) 
${E \over p_j}$ and ${1 \over p_j}$ have been expanded in powers of 
${1 \over E}$.

\end{document}